\documentclass[longauth]{aa}
\usepackage{graphicx}
\usepackage{txfonts}
\usepackage{xcolor}
\usepackage{hyperref}
\hypersetup{colorlinks, linkcolor=blue, citecolor=blue, urlcolor=blue}

\newcommand{\pivec}{\mbox{\boldmath $\pi$}}
\newcommand{\muvec}{\mbox{\boldmath $\mu$}}

\newcommand{\te}{t_{\rm E}}
\newcommand{\thetae}{\theta_{\rm E}}

\newcommand{\pie}{\pi_{\rm E}}
\newcommand{\pien}{\pi_{{\rm E},N}}
\newcommand{\piee}{\pi_{{\rm E},E}}
\newcommand{\dl}{D_{\rm L}}
\newcommand{\ds}{D_{\rm S}}

\definecolor{brown}{rgb}{0.59, 0.29, 0.0}
\definecolor{darkgreen}{rgb}{0.0, 0.42, 0.24}
\definecolor{darkblue}{rgb}{0.01, 0.31, 0.59}
\definecolor{darkblue}{rgb}{0.0, 0.25, 0.42}
\definecolor{blue}{rgb}{0.0,0.0,1.0}
\definecolor{green}{rgb}{0.0,1.0,0.0}

\begin{document}

\title{
Four binary microlenses with directly measured masses
}
\titlerunning{Four binary microlenses with directly measured masses}

\author{
     Cheongho~Han\inst{\ref{cbnu}}
\and Andrzej~Udalski\inst{\ref{warsaw}} 
\and Chung-Uk~Lee\inst{\ref{kasi}\thanks{\tt leecu@kasi.re.kr}} 
\and Ian~A.~Bond\inst{\ref{massey}}
\\
(Leading authors)
\\
     Michael~D.~Albrow\inst{\ref{canterbury}}   
\and Sun-Ju~Chung\inst{\ref{kasi}}      
\and Andrew~Gould\inst{\ref{osu}}      
\and Youn~Kil~Jung\inst{\ref{kasi},\ref{ust}} 
\and Kyu-Ha~Hwang\inst{\ref{kasi}} 
\and Yoon-Hyun~Ryu\inst{\ref{kasi}} 
\and Yossi~Shvartzvald\inst{\ref{weizmann}}   
\and In-Gu~Shin\inst{\ref{cfa}} 
\and Jennifer~C.~Yee\inst{\ref{cfa}}   
\and Weicheng~Zang\inst{\ref{cfa},\ref{tsinghua}}     
\and Hongjing~Yang\inst{\ref{tsinghua}}     
\and Sang-Mok~Cha\inst{\ref{kasi},\ref{kyunghee}} 
\and Doeon~Kim\inst{\ref{cbnu}}
\and Dong-Jin~Kim\inst{\ref{kasi}} 
\and Seung-Lee~Kim\inst{\ref{kasi}} 
\and Dong-Joo~Lee\inst{\ref{kasi}} 
\and Yongseok~Lee\inst{\ref{kasi},\ref{kyunghee}} 
\and Byeong-Gon~Park\inst{\ref{kasi}} 
\and Richard~W.~Pogge\inst{\ref{osu}}
\\
(The KMTNet Collaboration)
\\
     Przemek~Mr{\'o}z\inst{\ref{warsaw}} 
\and Micha{\l}~K.~Szyma{\'n}ski\inst{\ref{warsaw}}
\and Jan~Skowron\inst{\ref{warsaw}}
\and Rados{\l}aw~Poleski\inst{\ref{warsaw}} 
\and Igor~Soszy{\'n}ski\inst{\ref{warsaw}}
\and Pawe{\l}~Pietrukowicz\inst{\ref{warsaw}}
\and Szymon~Koz{\l}owski\inst{\ref{warsaw}} 
\and Krzysztof~A.~Rybicki\inst{\ref{warsaw},\ref{weizmann}}
\and Patryk~Iwanek\inst{\ref{warsaw}}
\and Krzysztof~Ulaczyk\inst{\ref{warwick}}
\and Marcin~Wrona\inst{\ref{warsaw},\ref{villanova}}
\and Mariusz~Gromadzki\inst{\ref{warsaw}}          
\and Mateusz~J.~Mr{\'o}z\inst{\ref{warsaw}} 
\and Micha{\l} Jaroszy{\'n}ski\inst{\ref{warsaw}}
\and Marcin Kiraga\inst{\ref{warsaw}}
\\
(The OGLE Collaboration)
\\
     Fumio~Abe\inst{\ref{nagoya}}
\and David~P.~Bennett\inst{\ref{nasa},\ref{maryland}}
\and Aparna~Bhattacharya\inst{\ref{nasa},\ref{maryland}}
\and Akihiko~Fukui\inst{\ref{tokyo-earth},}\inst{\ref{spain}}
\and Ryusei~Hamada\inst{\ref{osaka}}
\and Stela~Ishitani~Silva\inst{\ref{nasa},\ref{catholic}}  
\and Yuki~Hirao\inst{\ref{tokyo-ast}}
\and Naoki~Koshimoto\inst{\ref{tokyo-ast}}
\and Yutaka~Matsubara\inst{\ref{nagoya}}
\and Shota~Miyazaki\inst{\ref{osaka}}
\and Yasushi~Muraki\inst{\ref{nagoya}}
\and Tutumi~Nagai\inst{\ref{osaka}}
\and Kansuke~Nunota\inst{\ref{osaka}}
\and Greg~Olmschenk\inst{\ref{nasa}}
\and Cl{\'e}ment~Ranc\inst{\ref{sorbonne}}
\and Nicholas~J.~Rattenbury\inst{\ref{auckland}}
\and Yuki~Satoh\inst{\ref{osaka}}
\and Takahiro~Sumi\inst{\ref{osaka}}
\and Daisuke~Suzuki\inst{\ref{osaka}}
\and Sean K. Terry\inst{\ref{nasa}, \ref{maryland}}
\and Paul~J.~Tristram\inst{\ref{john}}
\and Aikaterini~Vandorou\inst{\ref{nasa},\ref{maryland}}
\and Hibiki~Yama\inst{\ref{osaka}}
\\
(The MOA Collaboration)
}

\institute{
      Department of Physics, Chungbuk National University, Cheongju 28644, Republic of Korea                                                          \label{cbnu}     
\and  Astronomical Observatory, University of Warsaw, Al.~Ujazdowskie 4, 00-478 Warszawa, Poland                                                      \label{warsaw}   
\and  Korea Astronomy and Space Science Institute, Daejon 34055, Republic of Korea                                                                    \label{kasi}   
\and  Institute of Natural and Mathematical Science, Massey University, Auckland 0745, New Zealand                                                    \label{massey}    
\and  University of Canterbury, Department of Physics and Astronomy, Private Bag 4800, Christchurch 8020, New Zealand                                 \label{canterbury}  
\and  Department of Astronomy, Ohio State University, 140 West 18th Ave., Columbus, OH 43210, USA                                                     \label{osu} 
\and  University of Science and Technology, Daejeon 34113, Republic of Korea                                                                          \label{ust}
\and  Department of Particle Physics and Astrophysics, Weizmann Institute of Science, Rehovot 76100, Israel                                           \label{weizmann}   
\and  Center for Astrophysics $|$ Harvard \& Smithsonian 60 Garden St., Cambridge, MA 02138, USA                                                      \label{cfa}  
\and  Department of Astronomy and Tsinghua Centre for Astrophysics, Tsinghua University, Beijing 100084, China                                        \label{tsinghua} 
\and  School of Space Research, Kyung Hee University, Yongin, Kyeonggi 17104, Republic of Korea                                                       \label{kyunghee}     
\and  Department of Physics, University of Warwick, Gibbet Hill Road, Coventry, CV4 7AL, UK                                                           \label{warwick}
\and  Villanova University, Department of Astrophysics and Planetary Sciences, 800 Lancaster Ave., Villanova, PA 19085, USA                           \label{villanova} 
\and  Institute for Space-Earth Environmental Research, Nagoya University, Nagoya 464-8601, Japan                                                     \label{nagoya}     
\and  Code 667, NASA Goddard Space Flight Center, Greenbelt, MD 20771, USA                                                                            \label{nasa} 
\and  Department of Astronomy, University of Maryland, College Park, MD 20742, USA                                                                    \label{maryland}  
\and  Department of Earth and Planetary Science, Graduate School of Science, The University of Tokyo, 7-3-1 Hongo, Bunkyo-ku, Tokyo 113-0033, Japan   \label{tokyo-earth} 
\and  Instituto de Astrof{\'i}sica de Canarias, V{\'i}a L{\'a}ctea s/n, E-38205 La Laguna, Tenerife, Spain                                            \label{spain} 
\and  Department of Earth and Space Science, Graduate School of Science, Osaka University, Toyonaka, Osaka 560-0043, Japan                            \label{osaka}  
\and  Department of Physics, The Catholic University of America, Washington, DC 20064, USA                                                            \label{catholic} 
\and  Institute of Astronomy, Graduate School of Science, The University of Tokyo, 2-21-1 Osawa, Mitaka, Tokyo 181-0015, Japan                        \label{tokyo-ast}
\and  Sorbonne Universit\'e, CNRS, UMR 7095, Institut d'Astrophysique de Paris, 98 bis bd Arago, 75014 Paris, France                                  \label{sorbonne}
\and  Department of Physics, University of Auckland, Private Bag 92019, Auckland, New Zealand                                                         \label{auckland}    
\and  University of Canterbury Mt.~John Observatory, P.O. Box 56, Lake Tekapo 8770, New Zealand                                                       \label{john}  
}                                                                                                                                                       
\date{Received ; accepted}

\abstract
{}
{
We investigated binary lens events from the 2022-2024 microlensing surveys, aiming to
identify events suitable for lens mass measurements. We focused on two key light curve
features: distinct caustic spikes with resolved crossings for measuring the angular Einstein
radius ($\theta_{\rm E}$), and long durations enabling microlens-parallax ($\pi_{\rm E}$)
measurements. Four events met these criteria: KMT-2022-BLG-1479, KMT-2023-BLG-0932,
OGLE-2024-BLG-0142, and KMT-2024-BLG-1309.
}
{
We estimated the angular Einstein radius by combining the normalized source radius measured
from modeling the resolved caustic spikes with the angular source radius derived from the
source color and magnitude. Additionally, we determined the microlens parallax through light
curve modeling, considering higher-order effects caused by the orbital motions of Earth and 
the binary lens.
}
{
With measurements of the event timescale, angular Einstein radius, and microlens parallax, we
uniquely determined the mass and distance of the lens. For the events KMT-2022-BLG-1479,
KMT-2023-BLG-0932, and KMT-2024-BLG-1309, both components of the binary lens have masses lower 
than that of the Sun, consistent with M-type dwarfs, which are the most common type of lenses 
in Galactic microlensing events. These lenses are relatively nearby, with distances $\lesssim 
2.5$ kpc, indicating their location within the Galactic disk.  In contrast, for OGLE-2024-BLG-0142, 
the primary lens component has a mass similar to that of the Sun, while the companion lens 
component has about half the mass of the primary. This lens system is situated at a greater 
distance, roughly 4.5 kpc.  
}
{}


\maketitle

\section{Introduction} \label{sec:one}

In most microlensing events involving a single-mass lens and a single source star (1L1S), 
the only observable that can be used to to infer the physical properties of the lens is 
the event timescale ($\te$). The timescale is related to the lens mass (\(M\)) and distance 
to the lens (\(\dl\)) through the following relations:  
\begin{equation}
\te = \frac{\thetae}{\mu}, \quad \thetae = \left( \kappa M \pi_{\rm rel} \right)^{1/2},  
\label{eq1}
\end{equation}  
where $\kappa = 4G/(c^2 {\rm AU}) \simeq 8.14~({\rm mas}/M_\odot)$, $\thetae$ is the angular 
Einstein radius,  $\mu$ is the relative lens-source proper motion, $\pi_{\rm rel} = {\rm AU}
(D_{\rm L}^{-1} - D_{\rm S}^{-1})$ denotes the relative lens-source parallax,  and $\ds$ is 
the source distance.  Because $\te$ depends on multiple parameters ($M$, $\dl$, and $\mu$), 
determining the lens mass and distance from the timescale alone is highly degenerate, making 
it challenging to uniquely constrain the lens properties.

For a subset of 1L1S lensing events, the angular Einstein radius can be measured 
as an additional observable. This measurement is feasible when the lens transits the surface 
of the source star, causing the magnification to reflect the intensity-weighted average across 
the source's surface. As a result, finite-source effects smooth out the peak of the lensing 
light curve \citep{Gould1994, Witt1994, Nemiroff1994}.  By modeling the light curve affected 
by finite-source effects, the normalized source radius,  $\rho = \theta_*/\thetae$, can be 
determined, where $\theta_*$ is the angular radius of the source star. Given that $\theta_*$ 
can be inferred from the source’s color and brightness, the angular Einstein radius is then 
obtained through  
\begin{equation}
\thetae = \frac{\theta_*}{\rho}.
\label{eq2}
\end{equation}

\begin{table*}[t]
\caption{Coordinates and ID correspondences.  \label{table:one}}
\begin{tabular}{llllllllllcc}
\hline\hline
\multicolumn{1}{c}{KMTNet}                           &
\multicolumn{1}{c}{$({\rm RA}, {\rm DEC})_{2000}$}      &
\multicolumn{1}{c}{$(l, b)$}                         &
\multicolumn{1}{c}{OGLE}                             &
\multicolumn{1}{c}{MOA}                              \\
\hline
 KMT-2022-BLG-1479   &  (17:53:23.69, -28:06:34.20)    &   $(1^\circ\hskip-2pt .5871, $-$1^\circ\hskip-2pt .0397)$   &    --                    &   MOA-2022-BLG-475   \\
 KMT-2023-BLG-0932   &  (18:15:11.73, -25:04:31.19)    &   $(6^\circ\hskip-2pt .6236, $-$3^\circ\hskip-2pt .8018)$   &    OGLE-2023-BLG-0756    &   MOA-2023-BLG-345   \\
 KMT-2024-BLG-0813   &  (18:10:34.71, -27:46:37.20)    &   $(3^\circ\hskip-2pt .7487, $-$4^\circ\hskip-2pt .1782)$   &    OGLE-2024-BLG-0142    &   MOA-2024-BLG-016   \\
 KMT-2024-BLG-1309   &  (18:04:10.72, -28:10:53.62)    &   $(2^\circ\hskip-2pt .7093, $-$3^\circ\hskip-2pt .1370)$   &    OGLE-2024-BLG-0761    &   MOA-2024-BLG-133   \\
\hline
\end{tabular}
\end{table*}

Another lensing observable that can provide an additional constraint on the physical 
parameters of the lens is the microlens parallax, defined as
\begin{equation}
\pie = {\pi_{\rm rel} \over \thetae}.
\label{eq3}
\end{equation}
Microlens parallax can be measured when the relative motion between the source and lens 
deviates from rectilinear due to the orbital motion of Earth around the Sun. This effect 
is known as ``annual microlens parallax'' \citep{Gould1992}. The resulting orbital 
acceleration introduces subtle asymmetries in the lensing light curve, causing deviations 
from the standard symmetric shape.  See Figure 1 of \citet{Gould2013}.  An alternative 
approach is to observe the event simultaneously from Earth and a space-based observatory, 
such as Spitzer or Kepler, which are separated from Earth by an AU-scale baseline. This 
``space-based microlens parallax'' method allows a direct measurement of $\pie$ by comparing 
the light curves observed from the two vantage points.  Measuring an additional lensing 
observable provides further constraints on the physical parameters of the lens, allowing 
for a more accurate determination of its properties.  When both additional observables 
$\thetae$ and $\pie$ are determined, the lens mass and distance can be uniquely determined 
using the relations from \citet{Gould2000}:  
\begin{equation}
M = \frac{\thetae}{\kappa \pie}, \quad \dl = \frac{\rm AU}{\pie \thetae + \pi_S},
\label{eq4}
\end{equation}
where $\pi_S$ denotes the parallax of the source.

In a 1L1S microlensing event, the probability of directly determining the lens mass by 
simultaneously measuring all relevant lensing observables is very low for the following 
reasons.  First, detecting finite-source effects in the light curve is only possible in 
a small subset of events where the lens-source impact parameter satisfies $u_0 < \rho$. 
In a typical Galactic microlensing event caused by a low-mass star, the angular Einstein 
radius is approximately $\thetae \sim 0.5$ mas. Given that a main-sequence source star 
in the bulge has an angular radius of about $\theta_* \sim 0.7 \times 10^{-3}$ mas, the 
normalized source radius is $\rho = \theta_*/\thetae \lesssim 10^{-3}$ for a main-sequence 
star and around $10^{-2}$ for a giant star. This implies that only a small fraction of 
events exhibit detectable finite-source effects.  Second, microlens parallax can be 
measured only in rare cases involving long-duration events that span a significant 
fraction of Earth orbital period.  As a result, determining lens masses by simultaneously 
measuring both $\thetae$ and $\pie$ using ground-based photometric data has been possible 
for only a very limited number of events, such as OGLE 2007-BLG-050L \citep{Batista2009} 
and MOA-2009-BLG-174L \citep{Choi2012}.

The likelihood of directly measuring the lens mass is significantly higher in events involving 
binary lenses. A binary-lens single-source (2L1S) event is typically identified by distinctive 
features in the lensing light curve that emerge when the source approaches or crosses a lensing 
caustic.  Caustics are regions in the source plane where the magnification of a point source 
becomes infinite \citep{Schneider1986, Mao1991}. These caustics form sets of one, two, or three 
closed curves, covering a substantial portion of the Einstein ring.  When a finite source passes 
near or across a caustic, higher-order derivatives of the magnification induce deviations from 
a point-source light curve \citep{Pejcha2009}.  This enables the measurement of $\rho$ and, 
consequently, the determination of $\thetae$. Moreover, the presence of two lenses increases 
the likelihood of a longer event timescale compared to a single-lens event. Additionally, 
well-resolved caustic features in the lensing light curve of a 2L1S event allow for precise 
measurement of subtle distortions caused by microlens parallax \citep{An2001}.

Although it remains challenging to fully determine the complete set of lensing observables, 
the physical parameters of the lens have been uniquely constrained for a subset of lensing 
events.  Binary-lens events with lens-mass determinations based on annual microlens parallax 
measurements include 
EROS~BLG-2000-5 \citep{An2002a}, OGLE 2003-BLG-267 \citep{Jaroszynski2005}, 
MOA-2009-BLG-016 \citep{Hwang2010}, OGLE-2009-BLG-020 \citep{Skowron2011}, MOA-2011-BLG-090 
and OGLE-2011-BLG-0417 \citep{Shin2012}, 
OGLE-2016-BLG-0156 \citep{Jung2019}, and 
KMT-2016-BLG-2052 \citep{Han2018}. 
Lens masses have also been determined for several binary-lens 
events through space-based microlens parallax using Spitzer, 
including OGLE-2014-BLG-1050 
\citep{Zhu2015}, OGLE-2019-BLG-0033 \citep{Herald2022}, OGLE-2016-BLG-1067 \citep{Calchi2019}, 
and OGLE-2017-BLG-1038 \citep{Malpas2022}.  For MOA-2015-BLG-020 \citep{Wang2017} and 
OGLE-2016-BLG-0168 \citep{Shin2017}, 
ground-based microlens parallax measurements were 
independently confirmed by Spitzer observations via the satellite parallax method.

In this study, we present direct lens mass measurements for four 2L1S events: KMT-2022-BLG-1479, 
KMT-2023-BLG-0932, OGLE-2024-BLG-0142, and KMT-2024-BLG-1309.  These events share common 
characteristics, including well-resolved caustic-crossing features in their light curves and 
long durations extending across a significant portion of a season or beyond. The well-resolved 
caustic features allowed for precise measurements of the angular Einstein radii, while the 
extended coverage of the lensing light curves enabled the determination of microlens parallax.  
By combining the measurements of $\thetae$ and $\pie$, we uniquely determined the lens masses 
for these events.

\section{Data} \label{sec:two}

We conducted a system analysis of 2L1S events detected by the Korea Microlensing Telescope 
Network \citep[KMTNet;][]{Kim2016} survey from the 2022 to 2024 seasons, aiming to identify 
events that allow for the measurement of lens masses. In selecting events, we focused on two 
key features in the lensing light curves. The first feature is the presence of distinct 
caustic spikes with well-resolved caustic crossings, which enable the measurement of the 
angular Einstein radius. The second feature is the long duration of the events, which span 
a significant portion of an observation season or even longer, facilitating the measurement 
of microlens parallax. Through this analysis, we identified four lensing events: 
KMT-2022-BLG-1479, KMT-2023-BLG-0932, KMT-2024-BLG-0813, and KMT-2024-BLG-1309.

We found that all the identified events were also observed by two other microlensing surveys: 
the Microlensing Observations in Astrophysics (MOA) \citep{Bond2001, Sumi2003} and the Optical 
Gravitational Lensing Experiment (OGLE) \citep{Udalski2015}.  Table~\ref{table:one} summarizes 
the event correspondences, showing the event IDs from the three surveys alongside their equatorial 
and Galactic coordinates.  The event KMT-2024-BLG-0813 was initially detected by the OGLE group. 
To align with the convention used in the microlensing community, we designate this event as 
OGLE-2024-BLG-0142, based on the ID assigned by the first discovery group. For our analysis, 
we use combined data from all three surveys.

The data for the microlensing events were obtained through observations made with the telescopes
operated by each survey group.  The KMTNet group has been conducting a lensing survey since 2015,
using three identical telescopes strategically placed in the Southern Hemisphere to ensure 
continuous monitoring of microlensing events.  These telescopes are located at Siding Spring 
Observatory in Australia (KMTA), Cerro Tololo Inter-American Observatory in Chile (KMTC), and 
South African Astronomical Observatory in South Africa (KMTS). Each KMTNet telescope has a 
1.6-meter aperture, and the camera mounted on each provides a 4 square-degree field of view.  
The OGLE group has been carrying out microlensing observations since 1992, with the current 
phase being its fourth.  The survey employs a 1.3-meter telescope, equipped with a camera 
offering a 1.4 square-degree field of view, located at Las Campanas Observatory in Chile.  The 
MOA survey began with a 0.6-meter telescope and is now in its second phase, using a 1.8-meter 
telescope at Mt. John University Observatory in New Zealand. The field of view of the camera is 
2.2 square degrees.  Observations by the KMTNet and OGLE surveys were conducted in the $I$ band, 
while the MOA survey observations were made in the customized MOA-$R$ band, covering a wavelength 
range of 609--1109 nm.

The light curves for the microlensing events were constructed using the photometric pipelines 
operated by each survey group: the KMTNet data were processed with the pipeline of 
\citet{Albrow2009}, the OGLE data were handled using the pipeline from \citet{Udalski2003}, 
and the MOA data were processed with the pipeline from \citet{Bond2001}.  For the KMTNet data, 
we used the reprocessed data obtained with the code developed by \citet{Yang2024} to ensure 
optimal data quality. Recognizing that pipeline-generated photometric errors often underestimate 
the true errors due to the omission of systematics, we recalibrated the error bars using the 
method described in \citet{Yee2012}.

\section{Lensing light curve analyses} \label{sec:three}

The light curves of all observed events displayed distinct caustic-crossing spikes, a 
hallmark of binary-lens events. To analyze these light curves, we modeled them using a 
2L1S configuration of the lens system. The modeling process aimed to identify a lensing 
solution that provides the best-fit lensing parameters for the observed light curve.

In the simplest case, where the relative motion between the lens and source is rectilinear, 
a 2L1S event is characterized by seven fundamental parameters. The first three, $(t_0, u_0, 
\te)$, describe the lens-source approach, where $t_0$ is the time of closest approach to a 
reference position on the lens plane, $u_0$ is the projected separation normalized to the 
Einstein radius $\thetae$ at that time, and $\te$ is the event timescale.  Two additional 
parameters, $(s, q)$, define the binary lens system, with $s$ representing the projected 
separation (scaled to $\thetae$) between the two lens components and $q$ denoting their mass 
ratio. The parameter $\alpha$ specifies the source trajectory angle relative to the binary-lens 
axis. Finally, $\rho$, the ratio of the angular source radius ($\theta_*$) to the angular 
Einstein radius, is required to describe caustic-crossing features influenced by finite-source 
effects.  The lens reference position is defined as the center of mass for close binaries with 
$s < 1.0$, and as the effective position described in \cite{An2002b} for wide binaries with 
$s > 1.0$.

In computing finite-source magnifications, we take into account the variation in surface 
brightness across the source star caused by limb darkening. Specifically, the surface-brightness 
profile is modeled as $S \propto 1 - \Gamma(1 - \frac{3}{2} \cos \phi)$, where $\Gamma$ 
represents a linear limb-darkening coefficient and $\phi$ is the angle between the line of 
sight to the observer and the local normal to the stellar surface \citep{Albrow1999}.  The 
limb-darkening coefficients are adopted from \citet{Claret2000}, based on the type of the 
source star, which is determined from its dereddened color and brightness as described in 
Sect.~\ref{sec:four}.

All events had extended durations, covering a substantial part of an observing season, and 
in some cases, extending into another season.  For events with long durations, two additional 
higher-order effects must be taken into account. The first is the microlens-parallax effect, 
which results from the acceleration of the relative lens-source motion due to the Earth’s 
orbital motion around the Sun \citep{Gould1992, Gould2000, Gould2004}.  The second is the 
lens-orbital effect, caused by the orbital motion of the binary lens \citep{Albrow2000, 
Batista2011, Skowron2011, Han2024}.  The orbital motion of the lens not only causes acceleration 
of the relative lens-source motion but also induces changes in the caustic structure.  The 
additional lensing parameters required to account for these higher-order effects are discussed 
below.

\begin{table*}[t]
\caption{Lensing parameters of KMT-2022-BLG-1479.\label{table:two}}
\begin{tabular}{lllllll}
\hline\hline
\multicolumn{1}{c}{Parameter}         &
\multicolumn{1}{c}{Static}            &
\multicolumn{2}{c}{Higher order}      \\
\multicolumn{1}{c}{         }         &
\multicolumn{1}{c}{      }            &
\multicolumn{1}{c}{$u_0>0$}           &
\multicolumn{1}{c}{$u_0<0$}      \\
\hline
 $\chi^2$                  &   $15134.1             $   &   $14917.1             $    &   $14920.5            $   \\
 $t_0$ (HJD$^\prime$)      &   $9828.6945 \pm 0.0087$   &   $9828.833 \pm 0.018  $    &   $9828.867 \pm 0.016 $   \\
 $u_0$                     &   $0.1791 \pm 0.0013   $   &   $0.2164 \pm 0.0041   $    &   $-0.2271 \pm 0.0044 $   \\
 $\te$ (days)              &   $50.86 \pm 0.20      $   &   $45.53 \pm 0.42      $    &   $44.34 \pm 0.41     $   \\
 $s$                       &   $0.7837 \pm 0.0017   $   &   $0.8301 \pm 0.0045   $    &   $0.8425 \pm 0.0047  $   \\
 $q$                       &   $0.6029 \pm 0.0037   $   &   $0.5955 \pm 0.0049   $    &   $0.5931 \pm 0.0051  $   \\
 $\alpha$ (rad)            &   $3.2032 \pm 0.0009   $   &   $3.2032 \pm 0.0021   $    &   $-3.2015 \pm 0.0022 $   \\
 $\rho$ ($10^{-3}$)        &   $6.379 \pm 0.043     $   &   $7.262 \pm 0.103     $    &   $7.546 \pm 0.116    $   \\
 $\pien$                   &   --                       &   $-0.246 \pm 0.080    $    &   $0.219 \pm 0.095    $   \\
 $\piee$                   &   --                       &   $0.0333 \pm 0.0083   $    &   $0.0645 \pm 0.0087  $   \\
 $ds/dt$ (yr$^{-1}$)       &   --                       &   $0.098 \pm 0.032     $    &   $0.093 \pm 0.034    $   \\
 $d\alpha/dt$ (yr$^{-1}$)  &   --                       &   $0.135 \pm 0.198     $    &   $-0.353 \pm 0.240   $   \\
\hline
\end{tabular}
\tablefoot{ ${\rm HJD}^\prime = {\rm HJD}- 2450000$.  }
\end{table*}

The modeling process began with the search for a static model that describes the overall light 
curve without incorporating higher-order effects. Given the complexity of exploring the entire 
parameter space due to the large number of lensing parameters, we adopted a hybrid approach that 
combines a grid search with a downhill optimization method. In this approach, the binary-lens 
parameters $(s, q)$ were explored using a grid search with multiple initial values for $\alpha$, 
while the remaining parameters were refined through a downhill optimization method. The grid 
search was conducted using the map-making technique developed by \citet{Dong2006}, while the 
downhill optimization was performed with a Markov chain Monte Carlo (MCMC) algorithm employing 
an adaptive step-size Gaussian sampler \citep{Doran2004}. This approach allowed us to investigate 
the $\chi^2$ distribution within the $(s, q, \alpha)$ parameter space to identify local minima 
that could indicate degeneracies in the lensing solution. For each identified local minimum, we 
further refined the solution by allowing all parameters to vary freely.

\begin{figure}[t]
\includegraphics[width=\columnwidth]{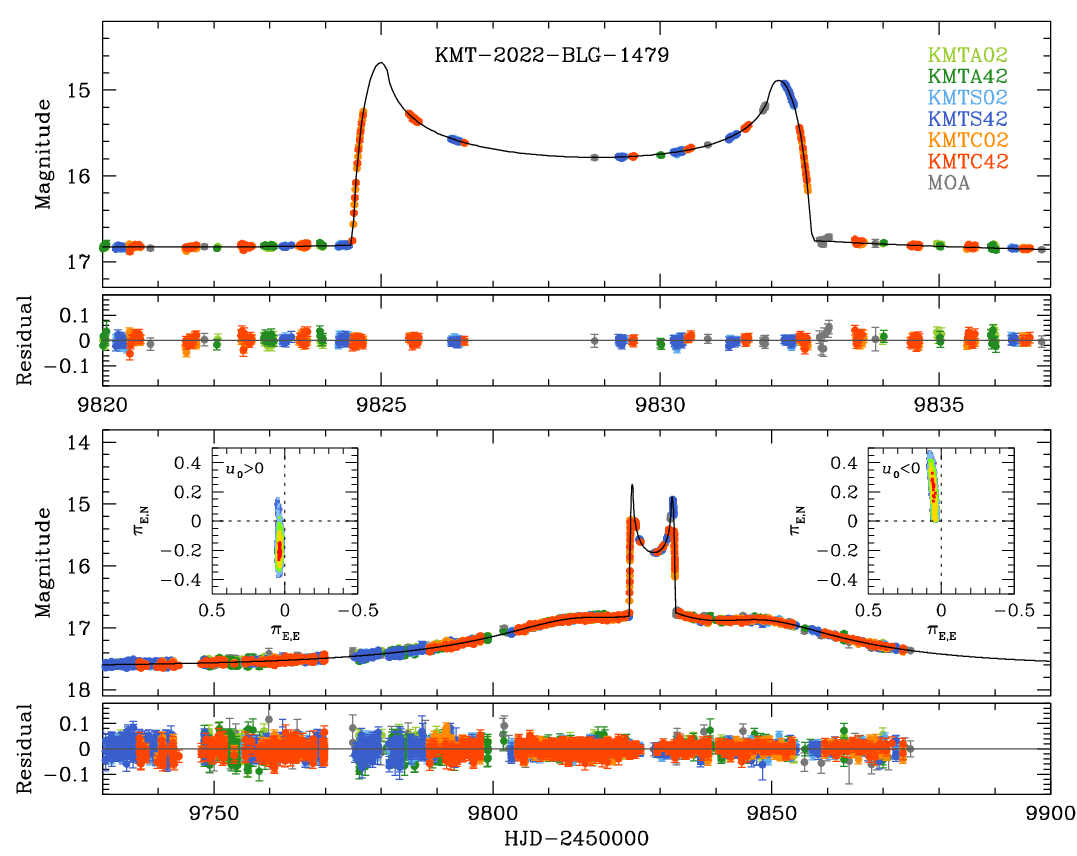}
\caption{
Light curve of the lensing event KMT-2022-BLG-1479. 
The two lower panels present the overall light curve and the residuals from the model, while 
the two upper panels provide a magnified view of the anomaly region. The curve overlaid on 
the data points represents the best-fit 2L1S model (higher-order model with $u_0>0$), which 
incorporates higher-order effects.  The colors of the data points correspond to the labels 
in the legend.  The two insets in the lower panel displays scatter plots of points from the 
MCMC chain in the $\piee$–$\pien$ parameter plane for the $u_0>0$ and $u_0<0$ solutions.  
Points within $1\sigma$, $2\sigma$, $3\sigma$, $4\sigma$, and $5\sigma$ confidence levels 
are marked in red, yellow, green, cyan, and blue, respectively.
}
\label{fig:one}
\end{figure}

The higher-order lensing parameters are determined based on the initial static solution. To 
achieve this, we first incorporated the lens-orbital effect and subsequently account for the 
additional microlens-parallax effect. Incorporating these higher-order effects necessitates 
the inclusion of additional parameters. To account for the microlens-parallax effect, the 
parameters $(\pien, \piee)$ are required, representing the north and east components of the 
microlens-parallax vector, $\pivec_{\rm E} = (\pi_{\rm rel}/\thetae) (\muvec/\mu)$.  Assuming 
that the positional changes of the lens components during lensing magnification are minimal, 
the lens-orbital effect is characterized by two parameters, ($ds/dt, d\alpha/dt$), which 
represent the rates of change in the binary lens separation and the source trajectory angle, 
respectively.  In the modeling, we imposed a constraint that the projected kinetic-to-potential 
energy ratio $({\rm KE}/{\rm PE})_\perp$ to be less than unity. This ratio is computed from the
model parameters as 
\begin{equation}
\left( {{\rm KE}\over {\rm PE}}\right)_\perp =
{(a_\perp/{\rm AU})^3 \over 8\pi^2(M/M_\odot)}
\left[
\left( {1\over s} {ds/dt\over {\rm yr}^{-1}}\right)^2+
\left( {d\alpha/dt \over {\rm yr}^{-1}}\right)^2
\right].
\label{eq5}
\end{equation}
Here $a_\perp$ represents the projected physical separation between the lens components. 
In the model that accounts for the microlens-parallax effect, we examined the ecliptic 
degeneracy between a pair of solutions with $u_0 > 0$ and $u_0 < 0$ \citep{Jiang2004, 
Poindexter2005}. For solutions exhibiting this degeneracy, the lensing parameters follow 
the relations $(u_0, \alpha, \pien, d\alpha/dt) \leftrightarrow -(u_0, \alpha, \pien, 
d\alpha/dt)$.  In the subsequent subsections, we detail the analyses of each individual 
event.

\begin{figure}[t]
\includegraphics[width=\columnwidth]{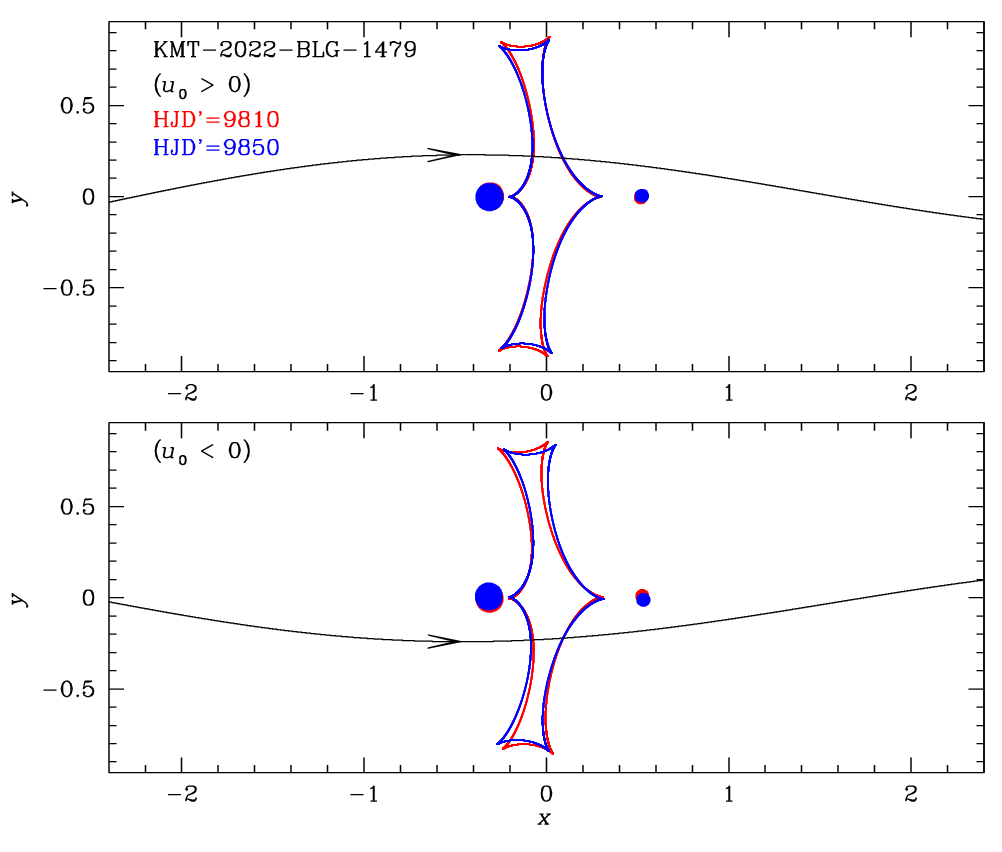}
\caption{
Lens system configuration of KMT-2022-BLG-1479
for the two higher-order models with $u_0>0$ and $u_0<0$.  In each panel, the trajectory 
of the source is indicated by the arrowed curve. The red and blue figures composed of 
concave curves represent the caustics at the times specified in the legend.  The positions 
of the lens components are marked by filled dots, with the larger dot corresponding to the 
more massive lens component.
}
\label{fig:two}
\end{figure}

\subsection{KMT-2022-BLG-1479} \label{sec:three-one}

The lensing event KMT-2022-BLG-1479 was initially detected by the KMTNet survey on July 11, 
2022 (${\rm HJD}^\prime \equiv {\rm HJD} - 2450000 = 9771$) and later identified by the MOA 
survey. The source of the event is located in the overlapping region of two KMTNet prime 
fields, BLG02 and BLG42, which were observed with cadences of 0.5 hours individually and 
0.25 hours when combined. The lensing magnification continued through the end of the 2022 
season, with data from the 2023 season also included in the analysis.

The light curve of KMT-2022-BLG-1479 is shown in Figure~\ref{fig:one}. It displays a pair 
of distinct caustic spikes that occurred at ${\rm HJD}^\prime \sim 9825$ and 9832, 
corresponding to the times when the source entered and exited a caustic, respectively. 
The spike at the caustic entrance was captured by the KMTC data set, while the spike at the 
caustic exit was resolved by the combination of the KMTC and KMTS data sets. In addition to 
these spikes, the light curve features two weak bumps: one before the first caustic spike, 
centered at ${\rm HJD}^\prime \sim 9810$, and the other after the second caustic spike, 
centered at ${\rm HJD}^\prime \sim 9850$. These bumps arose as the source approached the 
cusps of the caustic.

\begin{table*}[t]
\caption{Lensing parameters of KMT-2023-BLG-0932.\label{table:three}}
\begin{tabular}{lllllll}
\hline\hline
\multicolumn{1}{c}{Parameter}         &
\multicolumn{1}{c}{Static}            &
\multicolumn{2}{c}{Higher order}      \\
\multicolumn{1}{c}{         }         &
\multicolumn{1}{c}{      }            &
\multicolumn{1}{c}{$u_0>0$}           &
\multicolumn{1}{c}{$u_0<0$}           \\
\hline
 $\chi^2$                      &   $3746.6             $   &  $2276.4             $     &   $2280.4              $   \\
 $t_0$ (HJD$^{\prime\prime}$)  &   $192.6196 \pm 0.0053$   &  $193.1656 \pm 0.0372$     &   $193.1588 \pm 0.0335 $   \\
 $u_0$                         &   $0.05546 \pm 0.00021$   &  $0.04387 \pm 0.00185$     &   $-0.04398 \pm 0.00024$   \\
 $\te$ (days)                  &   $77.25 \pm 0.39     $   &  $124.90 \pm 6.00    $     &   $123.78 \pm 1.64     $   \\
 $s$                           &   $0.5869  \pm 0.0009 $   &  $0.5181  \pm 0.0093 $     &   $0.5187  \pm 0.0025  $   \\
 $q$                           &   $0.9555 \pm 0.0040  $   &  $0.710 \pm 0.025    $     &   $0.716 \pm 0.011     $   \\
 $\alpha$ (rad)                &   $4.7995 \pm 0.0011  $   &  $4.8149 \pm 0.0032  $     &   $-4.8118 \pm 0.0029  $   \\
 $\rho$ ($10^{-3}$)            &   $1.441 \pm 0.027    $   &  $0.967 \pm 0.047    $     &   $0.973 \pm 0.023     $   \\
 $\pien$                       &   --                      &  $-0.291 \pm 0.030   $     &   $0.303 \pm 0.0238    $   \\
 $\piee$                       &   --                      &  $0.0150 \pm 0.0084  $     &   $-0.0127 \pm 0.0045  $   \\
 $ds/dt$ (yr$^{-1}$)           &   --                      &  $-0.448 \pm 0.025   $     &   $-0.455 \pm 0.024    $   \\
 $d\alpha/dt$ (yr$^{-1}$)      &   --                      &  $-3.13 \pm 0.30     $     &   $3.14 \pm 0.13       $   \\
\hline
\end{tabular}
\tablefoot{ ${\rm HJD}^{\prime\prime} = {\rm HJD}- 2460000$.  }
\end{table*}

We began modeling the light curve using a static 2L1S configuration, which resulted in 
a unique solution with binary parameters $(s, q) \sim (0.78, 0.60)$ and an event timescale 
of $\te \sim 51$~days. Given the relatively long duration of the event, we examined whether 
incorporating higher-order effects would improve the fit. This led to a model that provided 
a significantly better fit than the static model, with an improvement of $\Delta\chi^2 = 
217.0$. In Table~\ref{table:one}, we present the model parameters for both the static and 
higher-order models, along with their respective $\chi^2$ values.  Between the two higher-order 
solutions, the one with $u_0 > 0$ provides a slightly better fit to the data.  The inclusion 
of higher-order effects results in a slight adjustment of the lensing parameters to $(s, q, \te) 
\sim (0.83, 0.60, 46 \text{ days})$. The model curve corresponding to the higher-order solution 
(with $u_0>0$) is plotted in Figure~\ref{fig:one}.  As shown in the scatter plots of MCMC points 
on the $(\piee, \pien)$ plane, presented in the insets of the lower panel of Figure~\ref{fig:one}, 
the microlens parallax was robustly measured as $(\pien, \piee) = (-0.246 \pm 0.080, 0.0333 \pm 
0.0083)$ for the $u_0>0$ solution and $(\pien, \piee) = (0.219 \pm 0.095, 0.0645 \pm 0.0087)$ 
for the $u_0<0$ solution.  Additionally, the normalized source radius was accurately determined 
from the well-resolved caustics, yielding $\rho = (7.262 \pm 0.103) \times 10^{-3}$ for the 
$u_0 > 0$ solution and $\rho = (7.546 \pm 0.116) \times 10^{-3}$ for the $u_0 < 0$ solution.

\begin{figure}[t]
\includegraphics[width=\columnwidth]{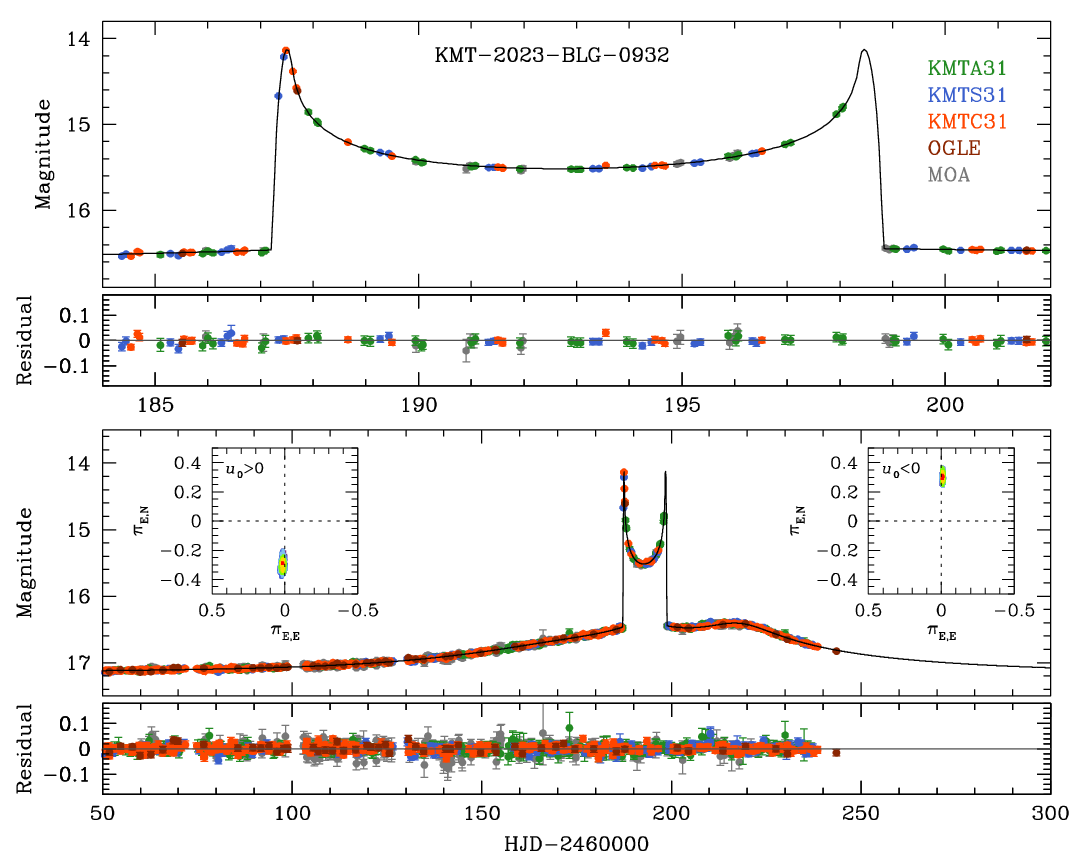}
\caption{
Light curve of the lensing event KMT-2023-BLG-0932. The notations used are consistent 
with those presented in Fig.~\ref{fig:one}. 
}
\label{fig:three}
\end{figure}

Figure~\ref{fig:two} illustrates the lens system configurations corresponding to the $u_0>0$ 
and $u_0<0$ solutions, depicting the source trajectory relative to the lens components and 
caustics.  The source trajectories of the two degenerate solutions are approximately symmetric 
with respect to the binary axis.  The curvature in the source trajectory arises because of 
the microlens-parallax effect. The positions of the lens components and the shape of the 
caustics vary due to the lens-orbital effect, with the lens positions and caustics marked 
at two specific epochs indicated in the legend.  The binary lens generates a single resonant 
caustic elongated in the direction nearly perpendicular to the binary lens axis.  The source 
moved almost parallel to the binary axis, crossing the caustic. These caustic crossings 
produced the sharp spike features in the lensing light curve. The weaker bumps observed 
before and after these spikes occurred as the source approached the left-side and right-side 
cusps of the caustic.

\subsection{KMT-2023-BLG-0932} \label{sec:three-two}

The microlensing event KMT-2023-BLG-0932 was first identified by the KMTNet group on May
28, 2023 (${\rm HJD}^{\prime\prime} \equiv {\rm HJD} - 2460000 = 92$), during the initial
phase of its magnification. It was later found by the MOA group on June 17 (${\rm
HJD}^{\prime\prime} = 112$) and by the OGLE group on July 16 (${\rm HJD}^{\prime\prime}
= 141$). Similar to the event KMT-2022-BLG-1479, the lensing-induced magnification persisted
throughout the 2023 season and continued until the end of the 2023 observing season. In the
analysis, we incorporate data from the 2024 season.  The source is located in the KMTNet 
BLG31 field, toward which observations were made with a 2.5-hour cadence.

Figure~\ref{fig:three} shows the lensing light curve of KMT-2023-BLG-0932. The light curve 
features two prominent spikes, occurring at ${\rm HJD}^{\prime\prime}\sim 187$ and 198, 
due to caustic crossings. While the second caustic crossing was not resolved, the first 
crossing was captured by the combined data from the three KMTNet telescopes and the OGLE 
telescope. In addition to these caustic spikes, a weak bump, caused by a cusp approach, 
appears around ${\rm HJD}^{\prime\prime}\sim 223$.

\begin{figure}[t]
\includegraphics[width=\columnwidth]{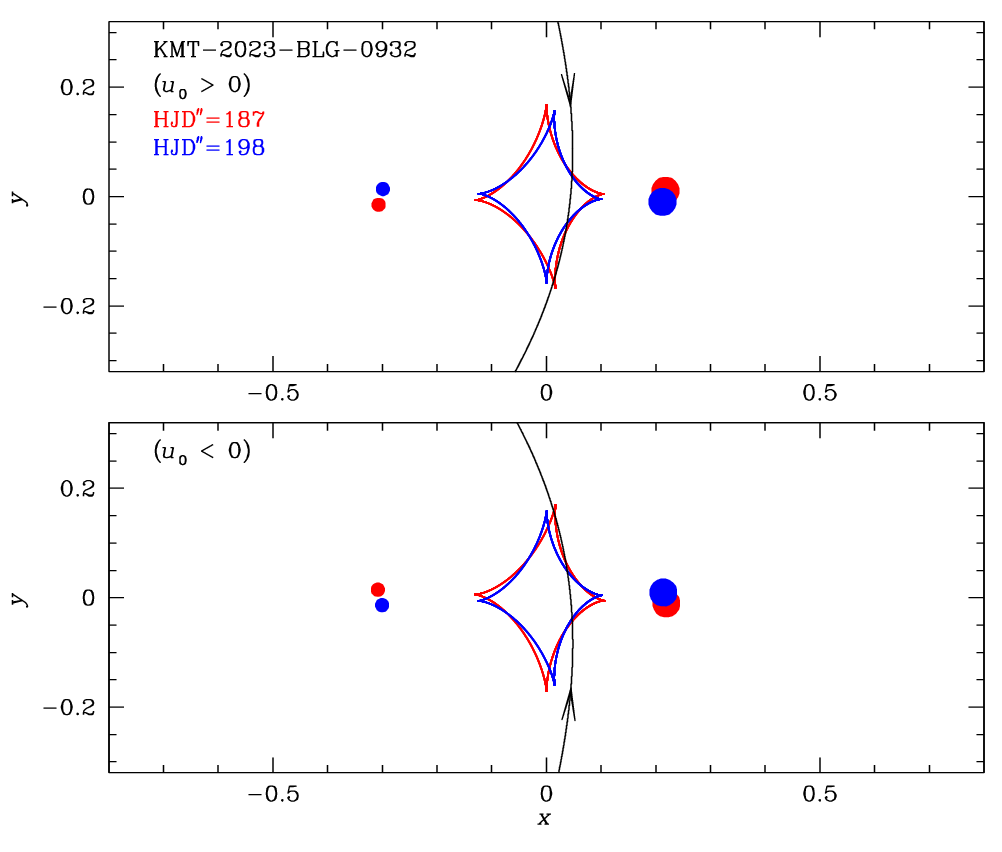}
\caption{
Lens system configuration of KMT-2023-BLG-0932. Notations are consistent with those 
in Fig.~\ref{fig:two}.
}
\label{fig:four}
\end{figure}

From the static 2L1S modeling, we obtained a solution with binary parameters $(s, q) \sim 
(0.59, 0.96)$ and an event timescale of $t_{\text{E}} \sim 77$ days.  However, this static 
model shows noticeable residuals.  When higher-order effects are included, these residuals 
are resolved, leading to a significant improvement in the fit.  The higher-order solution 
with $u_0 > 0$ provides a slightly better fit than the $u_0 < 0$ solution, with a difference 
of $\Delta\chi^2 = 4.0$.  The model curve of the $u_0>0$ solution  is overlaid on the data 
points, and the full set of lensing parameters for both the static and higher-order models 
are listed in Table~\ref{table:three}. The higher-order solution yields binary parameters 
of $(s, q) \sim (0.51, 1.41)$ and a timescale of $\te \sim 125$ days.  The insets of the 
lower panel displays scatter plots of MCMC points on the $(\piee, \pien)$ plane for the 
two degenerate higher-order solutions.  The normalized source radius was estimated from 
data points near the peak of the first spike, although its fractional uncertainty, $d\rho/
\rho \sim 4.9\%$, is larger than that of KMT-2022-BLG-1479 ($\sim 1.4\%$), for which both 
caustic crossings were densely covered.

\begin{table*}[t]
\caption{Lensing parameters of OGLE-2024-BLG-0142.\label{table:four}}
\begin{tabular}{lllllll}
\hline\hline
\multicolumn{1}{c}{Parameter}         &
\multicolumn{1}{c}{Static}            &
\multicolumn{2}{c}{Higher order}      \\
\multicolumn{1}{c}{         }         &
\multicolumn{1}{c}{      }            &
\multicolumn{1}{c}{$u_0>0$}           &
\multicolumn{1}{c}{$u_0<0$}           \\
\hline
 $\chi^2$                     &   $2551.7             $   &  $2500.1           $    &   $2498.4               $   \\
 $t_0$ (HJD$^{\prime\prime}$) &   $474.977 \pm 0.05414$   &  $474.652 \pm 0.086$    &   $474.6717 \pm 0.0856  $   \\
 $u_0$                        &   $0.1276 \pm 0.0019  $   &  $0.1213 \pm 0.0037$    &   $-0.1202 \pm 0.0023   $   \\
 $\te$ (days)                 &   $130.16 \pm 1.35    $   &  $140.02 \pm 3.87  $    &   $140.73 \pm 1.54      $   \\
 $s$                          &   $0.7343 \pm 0.0038  $   &  $0.7099 \pm 0.0084$    &   $0.7079 \pm 0.0019    $   \\
 $q$                          &   $0.554 \pm 0.006    $   &  $0.571 \pm 0.014  $    &   $0.576 \pm 0.010      $   \\
 $\alpha$ (rad)               &   $2.9246 \pm 0.0025  $   &  $2.9099 \pm 0.0065$    &   $-2.9083 \pm 0.0066   $   \\
 $\rho$ ($10^{-3}$)           &   $1.036 \pm 0.023    $   &  $0.99 \pm 0.031   $    &   $0.986 \pm 0.022      $   \\
 $\pien$                      &   --                      &  $-0.030 \pm 0.018 $    &   $0.059 \pm 0.0170     $   \\
 $\piee$                      &   --                      &  $0.066 \pm 0.013  $    &   $0.066 \pm 0.0137     $   \\
 $ds/dt$ (yr$^{-1}$)          &   --                      &  $0.120 \pm 0.026  $    &   $0.134 \pm 0.026      $   \\
 $d\alpha/dt$ (yr$^{-1}$   )  &   --                      &  $-0.101 \pm 0.041 $    &   $0.004 \pm 0.025      $   \\
\hline
\end{tabular}
\tablefoot{ ${\rm HJD}^{\prime\prime} = {\rm HJD}- 2460000$.  }
\end{table*}

Figure~\ref{fig:four} presents the lens system configurations for the higher-order solutions 
with $u_0 > 0$ and $u_0 < 0$.  In each panel, the two sets of caustics drawn in red and blue
correspond to the moments of caustic entrance and exit, respectively.  Since the mass ratio 
($q \sim 1.4$) is greater than one, the more massive lens component is located on the right 
side. In this event, the effect of lens orbital motion is significant, leading to a noticeable 
change in the caustic structure over the approximately 11-day interval between the two caustic 
crossings.  Recently, \citet{Han2024} reported three 2L1S events exhibiting lens-orbital effects: 
OGLE-2018-BLG-0971, MOA-2023-BLG-065, and OGLE-2023-BLG-0136. For KMT-2023-BLG-0932, the measured 
rate of change in the orientation angle, $\alpha \sim 3.1$/yr, is significantly larger than those 
observed in the events presented in that study.  The projected kinetic-to-potential energy ratio 
of the binary lens, calculated using Eq.~(\ref{eq5}), is $({\rm KE}/{\rm PE})_\perp \sim 0.48$.  
The binary lens generated three caustics: a single four-cusp central caustic, shown in the figure, 
and a pair of small three-cusp peripheral caustics, which are not depicted and are located away 
from the binary lens's barycenter.  The source approached the binary-lens axis at nearly a right 
angle, crossing the right side of the caustic, and then moved close to the cusp of the caustic. 
The caustic crossings generated the observed spikes, while the cusp approach produced the 
subsequent bump following the caustic spikes.

\begin{figure}[t]
\includegraphics[width=\columnwidth]{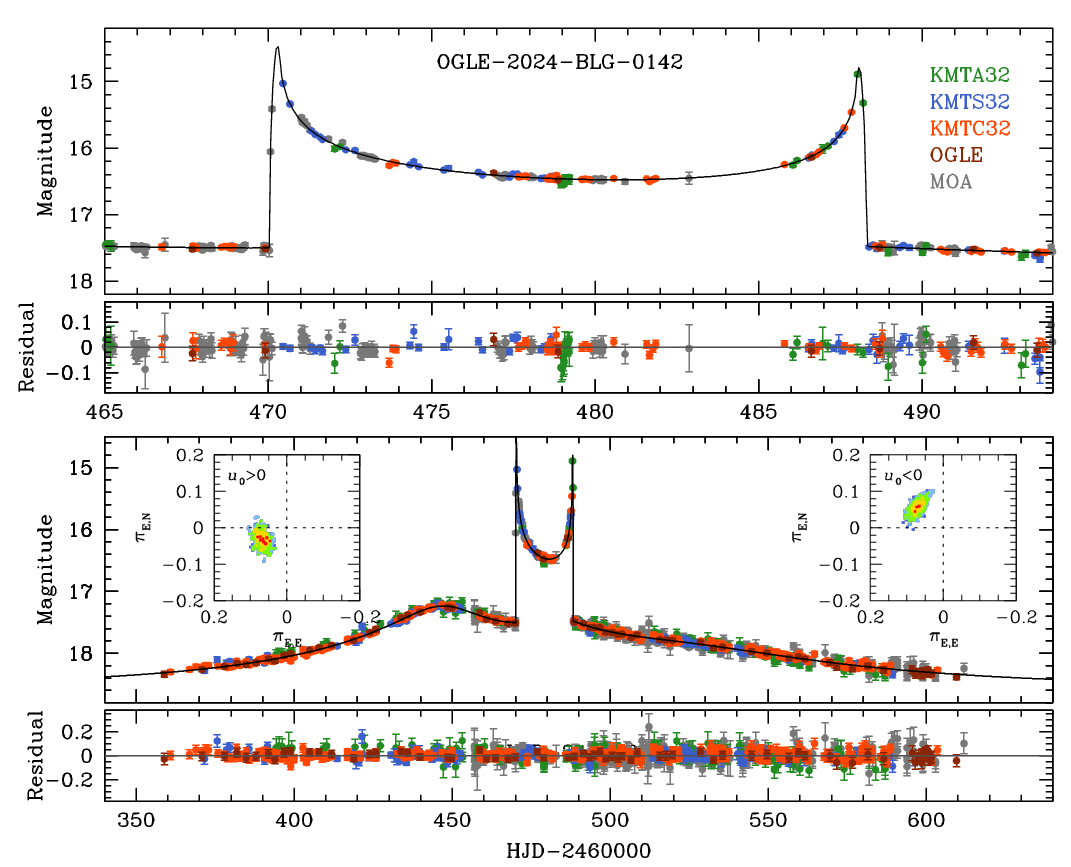}
\caption{
Lensing light curve of OGLE-2024-BLG-0142 and the best-fit model.
}
\label{fig:five}
\end{figure}

\subsection{OGLE-2024-BLG-0142} \label{sec:three-three}

The OGLE group initially detected the brightening of the source for the lensing event
OGLE-2024-BLG-0142 on March 20, 2024 (${\rm HJD}^{\prime\prime} = 389$) during its rising 
phase. The event was subsequently confirmed by the KMTNet group on May 3 (${\rm HJD}^{\prime
\prime} = 433$) and by the MOA group on May 28 (${\rm HJD}^{\prime\prime} = 458$). The 
magnification of the source flux began before the start of the 2024 season and continued 
throughout it. For our analysis, we included data from both the 2023 and 2025 seasons.

\begin{figure}[t]
\includegraphics[width=\columnwidth]{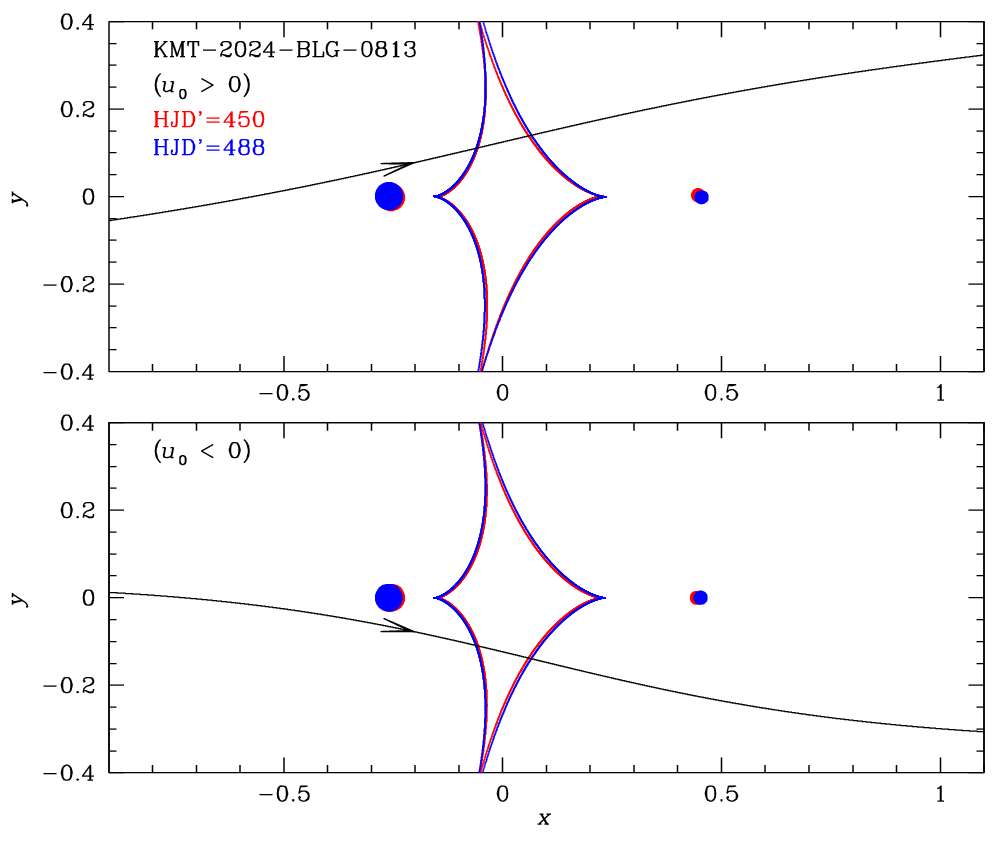}
\caption{
Lens-system configuration of OGLE-2024-BLG-0142. Notations are same as
those in Fig.~\ref{fig:two}.
}
\label{fig:six}
\end{figure}

The lensing light curve of the event is presented in Figure~\ref{fig:five}. Like the 
previous events, it features two prominent caustic spike features at ${\rm HJD}^{\prime
\prime} \sim 470$ and $\sim 488$. Both spikes were well-resolved, with the first captured 
by the combined data from KMTS and MOA, and the second by the combined data from KMTC and 
KMTS. In addition to these spikes, the light curve also shows a bump that appears before 
the first caustic spike, centered around ${\rm HJD}^{\prime\prime} \sim 445$.

We initially modeled the light curve under a static 2L1S configuration, and obtained 
a solution with binary parameters $(s, q) \sim (0.73, 0.55)$ and an event timescale of 
$\te \sim 130$ days. Given the long duration of the event, we further investigated a 
model incorporating microlens-parallax and lens-orbital effects, which resulted in a 
significantly improved fit with $\Delta\chi^2 = 51.6$. The best-fit parameters of the 
higher-order solution, $(s, q, \te) \sim (0.71, 0.57, 140)$, show slight variations 
from those of the static model.  Table~\ref{table:four} lists the complete sets of 
lensing parameters for the static and two higher-order models with $u_0>0$ and $u_0<0$.  
The $u_0 < 0$ model provides a slightly better fit than the \( u_0 > 0 \) solution, with 
an improvement of $\Delta\chi^2 = 1.7$.  Figure~\ref{fig:five} presents the model light 
curve of the $u_0<0$ higher-order solution along with its residuals, while the two insets 
in the lower panel displays the scatter plots of MCMC points in the $(\piee, \pien)$ 
parameter space for the two degenerate higher-order solutions.

\begin{figure}[t]
\includegraphics[width=\columnwidth]{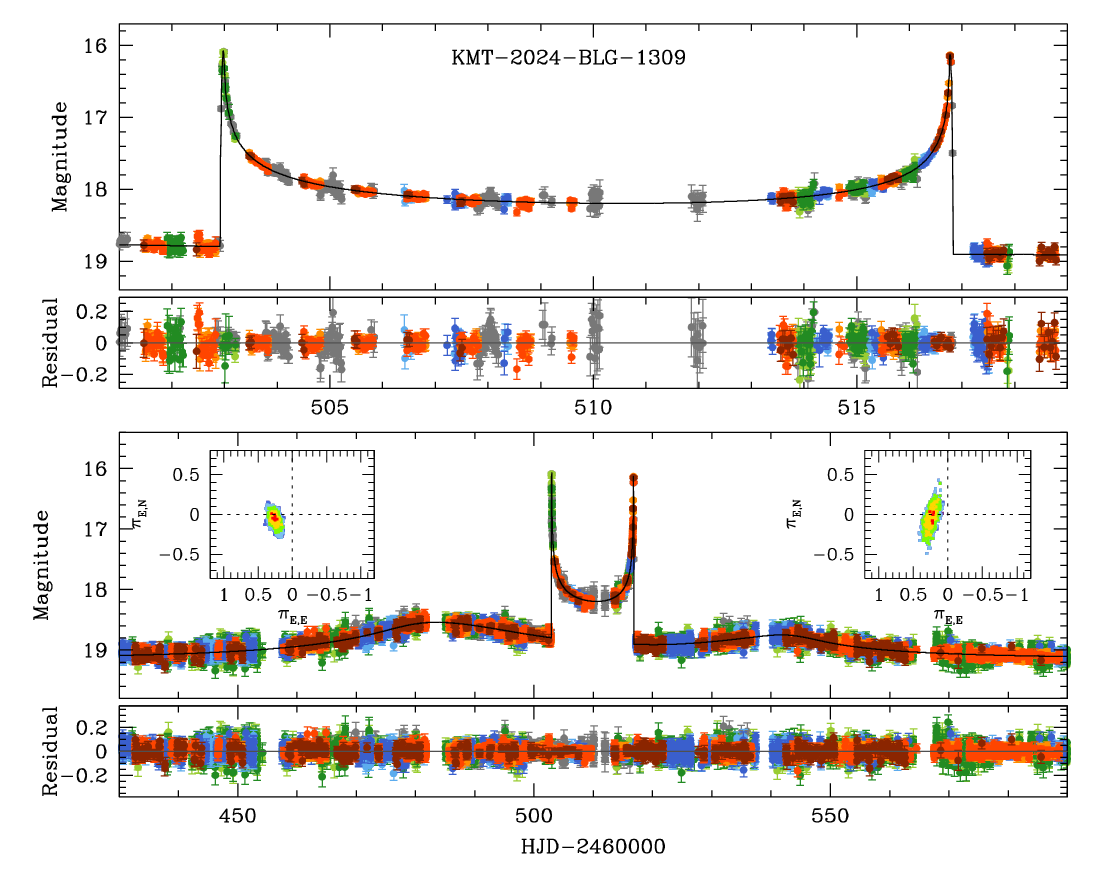}
\caption{
Lensing light curve of KMT-2024-BLG-1309 and the best-fit model.
}
\label{fig:seven}
\end{figure}

Figure~\ref{fig:six} illustrates the lens system configuration for the two higher-order 
solutions with $u_0>0$ and $u_0<0$.  Since the binary separation is smaller than the 
Einstein radius, the binary lens generates three caustics aligned perpendicular to the 
binary axis. The figure displays the central caustic, while the two peripheral caustics 
are not shown.  The source passed through the central caustic, producing the two 
caustic-induced spikes in the light curve. Before entering the caustic, the source 
approached the left-side on-axis cusp, generating a bump around ${\rm HJD}^{\prime\prime} 
\sim 445$.  We present two sets of caustics corresponding to the times of the bump and 
the second spike, separated by approximately 38 days. The influence of lens-orbital 
effects on the caustic structure is minimal, resulting in only slight changes during 
this period.

\begin{figure}[t]
\includegraphics[width=\columnwidth]{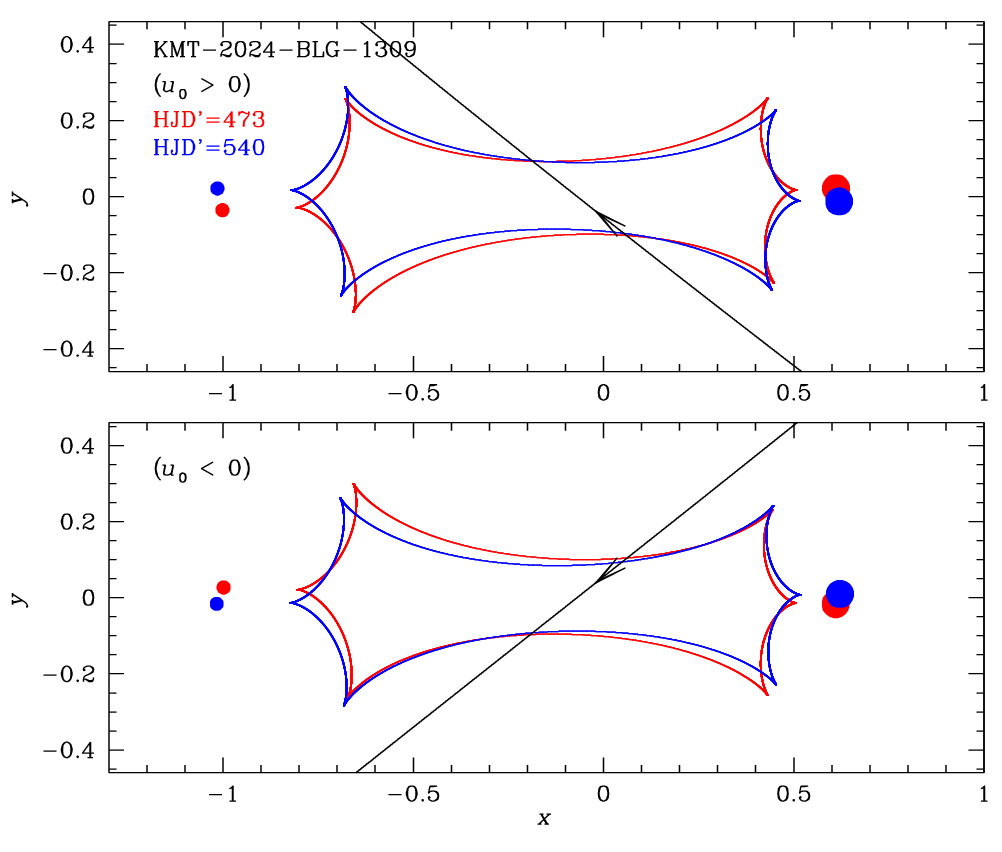}
\caption{
Lens system configuration of KMT-2024-BLG-1309.
}
\label{fig:eight}
\end{figure}

\begin{table*}[t]
\caption{Lensing parameters of KMT-2024-BLG-1309.\label{table:five}}
\begin{tabular}{lllllll}
\hline\hline
\multicolumn{1}{c}{Parameter}         &
\multicolumn{1}{c}{Static}            &
\multicolumn{2}{c}{Higher order}      \\
\multicolumn{1}{c}{         }         &
\multicolumn{1}{c}{      }            &
\multicolumn{1}{c}{$u_0>0$}           &
\multicolumn{1}{c}{$u_0<0$}           \\
\hline
 $\chi^2$                     &   $10046.7           $   &   $10009.0             $ &  $10008.5           $    \\
 $t_0$ (HJD$^{\prime\prime}$) &   $507.044 \pm 0.07  $   &   $507.71 \pm 0.06     $ &  $507.72 \pm 0.11   $    \\
 $u_0$                        &   $-0.0534 \pm 0.0012$   &   $  0.04296 \pm 0.0014$ &  $-0.0431 \pm 0.0020$    \\
 $\te$ (days)                 &   $44.190 \pm 0.081  $   &   $ 44.5023 \pm 0.111  $ &  $44.515 \pm 0.098  $    \\
 $s$                          &   $1.6181 \pm 0.0007 $   &   $  1.6264 \pm 0.0015 $ &  $1.6273 \pm 0.0018 $    \\
 $q$                          &   $0.6180 \pm 0.0003 $   &   $  0.6102 \pm 0.0048 $ &  $0.6112 \pm 0.0052 $    \\
 $\alpha$ (rad)               &   $5.6028 \pm 0.0014 $   &   $ -5.6147 \pm 0.0013 $ &  $5.6136 \pm 0.0055 $    \\
 $\rho$ ($10^{-3}$)           &   $0.4679 \pm 0.0056 $   &   $  0.4619 \pm 0.0062 $ &  $0.4695 \pm 0.0062 $    \\
 $\pien$                      &   --                     &   $ -0.046 \pm 0.067   $ &  $0.036 \pm 0.108   $    \\
 $\piee$                      &   --                     &   $  0.259 \pm 0.046   $ &  $0.261 \pm 0.054   $    \\
 $ds/dt$ (yr$^{-1}$)          &   --                     &   $  0.113 \pm 0.096   $ &  $0.157 \pm 0.093   $    \\
 $d\alpha/dt$ (yr$^{-1}$   )  &   --                     &   $ -0.3073 \pm 0.055  $ &  $0.234 \pm 0.102   $    \\
\hline
\end{tabular}
\tablefoot{ ${\rm HJD}^{\prime\prime} = {\rm HJD}- 2460000$.  }
\end{table*}

\begin{figure*}[t]
\centering
\sidecaption
\includegraphics[width=12.0cm]{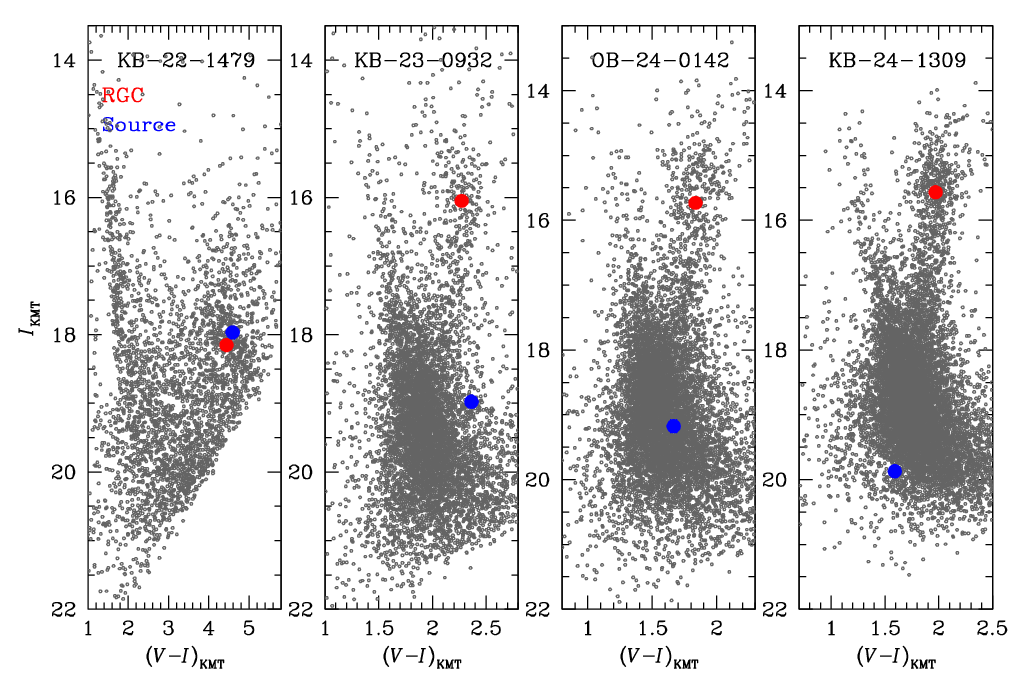}
\caption{
Source locations (blue dots) on the instrumental color-magnitude diagrams.
The centroids of the red giant clump (RGC) are also marked, serving as reference
points for color and magnitude calibration.
}
\label{fig:nine}
\end{figure*}

\subsection{KMT-2024-BLG-1309} \label{sec:three-four}

The microlensing event KMT-2024-BLG-1309 was observed by all three active survey groups. It 
was first detected by the KMTNet group on June 1, 2024 (${\rm HJD}^{\prime\prime}=472$), with 
subsequent detection by the OGLE group on June 17 (${\rm HJD}^{\prime\prime}=478$) and the 
MOA group on July 11 (${\rm HJD}^{\prime\prime}=502$). The source star has an $I$-band baseline 
magnitude of $I_{\rm base} = 19.16$ and is located in a field monitored with a high observational 
cadence. As a result, the event's light curve was densely sampled, providing detailed coverage 
of its magnification features.

Figure~\ref{fig:seven} shows the lensing light curve of KMT-2024-BLG-1309, which features two 
prominent caustic spikes at ${\rm HJD}^{\prime\prime} \sim 503.0$ and 516.8, along with two 
bumps at ${\rm HJD}^{\prime\prime} \sim 482$ before the first spike and at ${\rm HJD}^{\prime
\prime} \sim 538$ after the second caustic feature. Both caustic spikes were clearly resolved, 
with the first spike captured by the MOA and KMTS data sets, and the second spike observed by 
the OGLE, KMTC, and MOA data sets. Although the entire lensing magnification event occurred 
within the 2024 season, it lasted for a significant period of time.

We modeled the light curve under both a static 2L1S configuration and a configuration that 
accounts for higher-order effects. The lensing parameters derived from both models are similar, 
with values of $(s, q, \te) \sim (1.63, 1.64, 45~{\rm days})$. Incorporating higher-order 
effects leads to an improved fit, reducing the chi-square value by $\Delta\chi^2 = 38.2$. The 
scatter plot of the MCMC points on the $(\piee, \pien)$ plane for both solutions with $u_0>0$ 
and $u_0<0$ are presented in the inset of the lower panel of Figure~\ref{fig:seven}. A full 
sets of lensing parameters for both the static and higher-order models are provided in 
Table~\ref{table:five}. The model light curve obtained from the higher-order modeling (with 
$u_0>0$) is overlaid on the observational data.

Figure~\ref{fig:eight} shows the lens-system configurations corresponding to the $u_0 > 0$ 
(upper panel) and $u_0 < 0$ (lower panel) solutions.  The lens creates a single resonant 
caustic that extends along the binary-lens axis. The observed caustic spikes occurred as 
the source traversed diagonally across the central region of the caustic.  Before entering 
the caustic in the $u_0 < 0$ case, the source approached an upper cusp, and after exiting, 
it moved toward a lower cusp on the opposite side.  These cusp interactions lead to the 
formation of bumps in the light curve at the corresponding times.

\begin{table*}[t]
\caption{Source parameters, angular Einstein radii, and relative lens-source proper motions.  \label{table:six}}
\begin{tabular}{llllllllllcc}
\hline\hline
\multicolumn{1}{c}{Parameter}                &
\multicolumn{1}{c}{KMT-2022-BLG-1479}     &
\multicolumn{1}{c}{KMT-2023-BLG-0932}     &
\multicolumn{1}{c}{OGLE-2024-BLG-0142}    &
\multicolumn{1}{c}{KMT-2024-BLG-1309}     \\
\hline
 $(V-I)_s$                  &  $4.593 \pm 0.043 $    &   $2.364 \pm 0.009  $   &  $1.667 \pm 0.024 $    &  $1.591 \pm 0.008 $    \\
 $I_s$                      &  $17.966 \pm 0.001$    &   $18.976 \pm 0.001 $   &  $19.176 \pm 0.002$    &  $19.872 \pm 0.003$    \\
 $(V-I, I)_{\rm RGC}$       &   (4.442, 18.150)      &     (2.273, 16.050)     &    (1.836, 15.731)     &    (1.972, 15.568)     \\
 $(V-I, I)_{{\rm RGC},0}$   &   (1.060, 14.382)      &     (1.060, 14.257)     &    (1.060, 14.334)     &    (1.060, 14.357)     \\
 $(V-I)_{s,0}$              &  $1.211 \pm 0.059 $    &   $1.151 \pm 0.041  $   &  $0.891 \pm 0.047 $    &  $0.679 \pm 0.041 $    \\
 $I_{s,0}$                  &  $14.199 \pm 0.020$    &   $17.183 \pm 0.020 $   &  $17.780 \pm 0.020$    &  $18.660 \pm 0.020$    \\
 Spectral type              &   K4III                &     K4IV                &    K1.5V               &    G1V                 \\
 $\theta_*$ ($\mu$as)       &  $7.78 \pm 0.71   $    &   $1.88 \pm 0.17    $   &  $1.071 \pm 0.090 $    &  $0.562 \pm 0.046 $    \\
 $\thetae$ (mas)            &  $1.07 \pm 0.10   $    &   $1.89 \pm 0.19    $   &  $1.081 \pm 0.172 $    &  $1.197 \pm 0.101 $    \\
 $\mu$ (mas/yr)             &  $8.60 \pm 0.80   $    &   $5.70 \pm 0.59    $   &  $2.82 \pm 0.45   $    &  $9.82 \pm 0.83   $    \\
\hline                                                             
\end{tabular}                                                      
\end{table*}

\section{Source stars and angular Einstein radii} \label{sec:four}

In this section, we specify the source stars to fully characterize the events and estimate the 
angular Einstein radii. The source of each event is defined by measuring its $V-I$ color and 
$I$-band magnitude. After calibrating these values using the procedure outlined below, we 
estimate the angular source radius ($\theta_*$) based on the calibrated color and magnitude. 
With the estimated $\theta_*$, we calculate the angular Einstein radius using the relationship 
in Equation~(\ref{eq2}), along with the normalized source radius $\rho$ derived from light 
curve modeling. The following provides a detailed, step-by-step description of the procedure 
for measuring $\thetae$.

In the first step, we measured the instrumental source magnitudes in the $V$ and $I$ 
passbands.  To do this, we created two light curves from the data measured in these passbands, 
which were processed using the pyDIA code developed by \citet{Albrow2017}. From these light 
curves, we determined the instrumental flux of the source ($F_s$) in each band by fitting 
the observed flux ($F_{\rm obs}$) to the model ($A_{\rm model}$);
\begin{equation}
F_{\rm obs} =A_{\rm model}(t) F_s + F_b.
\label{eq6}
\end{equation}
Here, $F_b$ represents the blended flux. Figure~\ref{fig:nine} illustrates the source 
locations in the instrumental color-magnitude diagrams (CMDs), which were constructed 
using PYDIA photometry of stars located near the source.

The source color and magnitude were calibrated based on their position in the CMD. For this 
calibration, the centroid of the red giant clump (RGC) in the CMD was used as a reference 
\citep{Yoo2004}, as its de-reddened color and magnitude, $(V-I, I)_{{\rm RGC},0}$, corrected 
for extinction, had been previously determined by \citet{Bensby2013} and \citet{Nataf2013}. 
By measuring the color and magnitude offset, $\Delta(V-I, I)$, between the source and the RGC 
centroid, the de-reddened source color and magnitude were calculated as
\begin{equation}
(V-I, I)_{s,0} = (V-I, I)_{{\rm RGC},0} + \Delta(V-I, I).
\label{eq7}
\end{equation}
Table~\ref{table:six} lists the instrumental color and magnitude of the source, $(V-I, I)_s$, 
and the RGC centroid, $(V-I, I)_{\rm RGC}$, along with the de-reddened values for both the RGC, 
$(V-I)_{{\rm RGC},0}$, and the source, $(V-I, I)_{s,0}$, for each event. Based on the estimated 
colors and magnitudes, the source stars are classified as follows: a K-type giant for 
KMT-2022-BLG-1479, a K-type subgiant for KMT-2023-BLG-0932, an early K-type main-sequence star 
for OGLE-2024-BLG-0142, and an early G-type main-sequence star for KMT-2024-BLG-1309.

The angular radius of the source star was determined using the estimated color and magnitude.
This was achieved by applying the relation between $\theta_*$ and $(V-K, K)$ provided by
\citet{Kervella2004}. To utilize this relation, the $V-I$ color was converted to $V-K$ using 
the color-color relation from \citet{Bessell1988}, enabling the derivation of $\theta_*$. With 
the angular source radius estimated, the angular Einstein radius, $\thetae$, was calculated 
using the relation in Equation~(\ref{eq2}). Combined with the event timescale obtained from 
the modeling, the relative lens-source proper motion, $\mu$, was determined as
\begin{equation}
\mu = \frac{\thetae}{\te}.
\label{eq8}
\end{equation}
The derived values of $\theta_*$, $\thetae$, and $\mu$ for the events are provided in the 
last three rows of Table~\ref{table:six}.

\begin{table*}[t]
\footnotesize
\caption{Physical lens parameters.  \label{table:seven}}
\begin{tabular}{llllllllllcc}
\hline\hline
\multicolumn{1}{c}{Parameter}             &
\multicolumn{2}{c}{KMT-2022-BLG-1479}     &
\multicolumn{2}{c}{KMT-2023-BLG-0932}     &
\multicolumn{2}{c}{OGLE-2024-BLG-0142}    &
\multicolumn{2}{c}{KMT-2024-BLG-1309}     \\
\multicolumn{1}{c}{ }             &
\multicolumn{1}{c}{($u_0>0$)}     &
\multicolumn{1}{c}{($u_0<0$)}     &
\multicolumn{1}{c}{($u_0>0$)}     &
\multicolumn{1}{c}{($u_0<0$)}     &
\multicolumn{1}{c}{($u_0>0$)}     &
\multicolumn{1}{c}{($u_0<0$)}     &
\multicolumn{1}{c}{($u_0>0$)}     &
\multicolumn{1}{c}{($u_0<0$)}     \\
\hline             
 $M_1$ ($M_\odot$) &  $0.33 \pm 0.11$  & $0.35 \pm 0.12$ &  $0.343 \pm 0.048$  &  $0.342 \pm 0.048$   &  $1.17 \pm 0.35$  &  $0.94 \pm 0.26$   &  $0.214 \pm 0.059$  &  $0.211 \pm 0.057$   \\
 $M_2$ ($M_\odot$) &  $0.20 \pm 0.07$  & $0.21 \pm 0.07$ &  $0.471 \pm 0.066$  &  $0.481 \pm 0.067$   &  $0.67 \pm 0.20$  &  $0.55 \pm 0.15$   &  $0.352 \pm 0.097$  &  $0.346 \pm 0.092$   \\
 $\dl$ (kpc)       &  $2.54 \pm 0.60$  & $2.76 \pm 0.66$ &  $1.48 \pm 0.17  $  &  $1.42 \pm 0.17  $   &  $4.76 \pm 0.71$  &  $4.38 \pm 0.65$   &  $2.22 \pm 0.45  $  &  $2.24 \pm 0.44  $   \\
 $a_\perp$ (AU)    &  $2.26 \pm 0.53$  & $2.40 \pm 0.57$ &  $1.46 \pm 0.17  $  &  $1.44 \pm 0.17  $   &  $3.65 \pm 0.54$  &  $3.36 \pm 0.50$   &  $4.39 \pm 0.89  $  &  $4.38 \pm 0.85  $   \\
\hline
\end{tabular}
\end{table*}

\section{Physical lens parameters} \label{sec:five}

Using the lensing observables $\te$, $\thetae$, and $\pie$, we derived the physical parameters 
of the lens mass and distance based on the relationships described in Equations~(\ref{eq4}). 
The determined physical lens parameters, including the masses of the binary lens components 
($M_1$ and $M_2$), the lens distance ($\dl$), and the projected separation between the lens 
components ($a_\perp$), corresponding to the $u_0>0$ and $u_0<0$ solutions are presented in 
Table~\ref{table:seven}. The projected separation is calculated using the equation: 
\begin{equation} 
a_\perp = s \dl \thetae, 
\label{eq9}
\end{equation} 
where the normalized source radius $s$ is obtained from the modeling process. For the estimation 
of the physical parameters, we assume the distance to the source to be $\ds = 8$ kpc.  Note, 
however, that the mass estimates, $M=\thetae/(\kappa \pie)$, are independent of the assumed 
source distances.  Although some variations exist, the physical parameters estimated from the 
$u_0>0$ and $u_0<0$ solutions remain consistent.

The masses of the lens components vary across different events. In the cases of KMT-2022-BLG-1479, 
KMT-2023-BLG-0932, and KMT-2024-BLG-1309, both components of the binary lens are less massive than 
the Sun, with their masses aligning with those of M-type dwarfs, which are the most common type of 
lenses in Galactic microlensing events \citep{Han2003}.  In contrast, for OGLE-2024-BLG-0142, the 
primary lens component has a mass comparable to that of the Sun, while the companion lens component 
has a mass approximately half that of the primary.

The distances to the lenses also vary across different events. For KMT-2022-BLG-1479, KMT-2023-BLG-0932, 
and KMT-2024-BLG-1309, the lenses are located at relatively close distances, with $\dl \lesssim 2.7$ 
kpc, suggesting that these lenses are situated within the Galactic disk.  In contrast, the lens for 
OGLE-2024-BLG-0142 is located farther away, with a distance of $\dl \sim 4.8$ kpc for the $u_0 > 0$ 
solution and $\dl \sim 4.4$ kpc for the $u_0 < 0$ solution, which is significantly greater than that 
of the other events.  Bayesian analysis, incorporating Galactic model priors, indicates that the lens 
for OGLE-2024-BLG-0142 has a 35\% probability of being in the disk and a 65\% probability of being 
in the bulge.

\section{Summary and conclusion}  \label{sec:six}

We analyzed binary lens events detected from microlensing surveys from 2022 to 2024 seasons to
identify those suitable for lens mass measurements. Our investigation focused on two key features
in lensing light curves. The first feature is distinct caustic spikes with resolved crossings, 
which allow for the measurement of the angular Einstein radius $\theta_{\rm E}$. The second 
feature is long event durations, which enable the determination of the microlens parallax 
$\pi_{\rm E}$.  Based on these criteria, we identified four qualifying events, 
KMT-2022-BLG-1479, KMT-2023-BLG-0932, OGLE-2024-BLG-0142, and KMT-2024-BLG-1309.

Detailed modeling of the events revealed that the lenses are binary systems with mass ratios
between 0.5 and 0.7.  For all events, in addition to the basic event timescale, the extra 
observables of the angular Einstein radius and microlens parallax were securely measured. 
The angular Einstein radius was derived by combining the normalized source radius obtained 
from modeling the resolved caustic spikes with the angular source radius estimated from the 
source color and magnitude. The microlens parallax was determined through light curve modeling, 
accounting for higher-order effects induced by the orbital motions of Earth and the binary lens.

By combining the event timescale, angular Einstein radius, and microlens parallax, the mass 
and distance of lens were uniquely determined. In the cases of KMT-2022-BLG-1479, KMT-2023-BLG-0932, 
and KMT-2024-BLG-1309, both components of the binary lens have masses below that of the Sun, aligning 
with M-type dwarfs, which are the most frequently observed lenses in Galactic microlensing events. 
These lenses are located relatively close, at distances below 2.7 kpc, suggesting their presence 
within the Galactic disk.  For OGLE-2024-BLG-0142, the primary lens component has a mass similar 
to that of the Sun, while the companion lens component has about half the mass of the primary.  
This lens system is located at a greater distance, approximately 4.4 kpc for one solution and 
4.8 kpc for the other.

\begin{table}[t]
\caption{Previous events with physical lens parameters.  \label{table:eight}}
\begin{tabular*}{\columnwidth}{@{\extracolsep{\fill}}lllcc}
\hline\hline
\multicolumn{1}{c}{Event              }     &
\multicolumn{1}{c}{$M_1$ ($M_\odot$)  }     &
\multicolumn{1}{c}{$M_2$ ($M_\odot$)  }     &
\multicolumn{1}{c}{$D_{\rm L}$ (kpc)  }     \\
\hline             
EROS~BLG-2000-5$^{(1)}$      &  0.35    &  0.26                    &  2.6 \\
OGLE 2003-BLG-267$^{(2)}$    &  0.052   &  0.065                   &  5.4 \\
MOA-2009-BLG-016$^{(3)}$     &  0.13    &  0.04                    &  4.7 \\
OGLE-2009-BLG-020$^{(4)}$    &  0.84    &  0.23                    &  1.1 \\
MOA-2011-BLG-090$^{(5)}$     &  0.43    &  0.39                    &  3.3 \\
OGLE-2011-BLG-0417$^{(5)}$   &  0.57    &  0.17                    &  0.9 \\
OGLE-2016-BLG-0156$^{(6)}$   &  0.18    &  0.16                    &  1.4 \\
KMT-2016-BLG-2052$^{(7)}$    &  0.34    &  0.17                    &  2.1 \\
\hline
\end{tabular*}
\tablefoot{ 
(1) \citet{An2002b},
(2) \citet{Jaroszynski2005},
(3) \citet{Hwang2010},
(4) \citet{Skowron2011},
(5) \citet{Shin2012},
(6) \citet{Jung2019}, and
(7) \citet{Han2018}.
}
\end{table}

For a comparison between the physical parameters of the current events and those of previously 
reported events with measured annual parallaxes (as discussed in Sect.  \ref{sec:one}), we 
present their respective lens masses and distances in Table~\ref{table:eight}.  From this 
comparison, two notable trends emerge.  First, the lens masses are generally not large, despite 
the long timescales observed in the events. This indicates that the extended durations are 
primarily due to a combination of slow lens-source proper motions and large angular Einstein 
radii, rather than being driven by massive lenses, as previously mentioned by \citet{Han2018}.  
Second, most lenses are located at relatively close distances, typically within $\dl \lesssim 
5$ kpc.  This suggests that they predominantly reside in the Galactic disk, although, for the 
overall microlensing event population, the contribution from bulge lenses generally exceeds 
that of disk lenses \citep{Han2003}. This tendency is expected, because nearby lenses generally 
exhibit larger microlens parallaxes, which facilitates their characterization.

The binary-lens events analyzed in this study suggest that the upcoming gravitational microlensing 
experiment using the Nancy Grace Roman Space Telescope \citep{Spergel2015} will allow for the 
determination of lens masses in many binary-lens events. While the primary focus of the Roman 
microlensing survey is the detection of distant exoplanets, including those with masses smaller 
than Earth's, it will also uncover numerous binary-lens events with densely resolved, high-precision 
light curves, made possible by the telescope's excellent photometric accuracy and 15-minute cadence 
of space-based observations. For many of these events, as shown in this study, the physical properties 
of the lens are expected to be directly measurable.

\begin{acknowledgements}
This research was supported by the Korea Astronomy and Space Science Institute under the R\&D 
program (Project No. 2025-1-830-05) supervised by the Ministry of Science and ICT.
This research has made use of the KMTNet system operated by the Korea Astronomy and Space Science 
Institute (KASI) at three host sites of CTIO in Chile, SAAO in South Africa, and SSO in Australia. 
Data transfer from the host site to KASI was supported by the Korea Research Environment Open NETwork 
(KREONET). 
C.Han acknowledge the support from the Korea Astronomy and Space Science Institute under the R\&D 
program (Project No. 2025-1-830-05) supervised by the Ministry of Science and ICT.
J.C.Y., I.G.S., and S.J.C. acknowledge support from NSF Grant No. AST-2108414. 
W.Zang acknowledges the support from the Harvard-Smithsonian Center for Astrophysics 
through the CfA Fellowship.
The MOA project is supported by JSPS KAKENHI Grant Number 
JP16H06287, JP22H00153 and 23KK0060.
C.R. was supported by the Research fellowship of the Alexander von Humboldt Foundation.
\end{acknowledgements}

\end{document}